\documentclass[journal=jpccck,manuscript=article]{achemso}
\usepackage{amsmath,color}
\usepackage{graphicx}
\usepackage{amssymb}
\usepackage{caption}
\usepackage{subcaption}
\usepackage{rotating}
\usepackage[version=3]{mhchem}

\author{Sandeep K. Jain}
\affiliation[Institute for Theoretical Physics,Universiteit Utrecht]
{Institute for Theoretical Physics, Universiteit Utrecht, Leuvenlaan 4, 3584 CE Utrecht, The Netherlands}
\email{S.K.Jain@uu.nl}
\author{Gerard T. Barkema}
\affiliation[Institute for Theoretical Physics,Universiteit Utrecht] 
{Institute for Theoretical Physics, Universiteit Utrecht, Leuvenlaan 4, 3584 CE Utrecht, The Netherlands}
\alsoaffiliation[Universiteit Leiden]
{Institut-Lorentz, Universiteit Leiden, Niels Bohrweg 2, 2333 CA Leiden, The Netherlands}
\author{Normand Mousseau}
\affiliation[Universit\'{e} de Montr\'{e}al]
{D\'{e}partement de Physique and Regroupement Qu\'{e}b\'{e}cois sur les Mat\'{e}riaux de Pointe, Universit\'{e} de Montr\'{e}al, Case Postale 6128, succ. Centre-ville, Montr\'{e}al, Qu\'{e}bec H3C 3J7, Canada}
\author{Chang-Ming Fang}
\affiliation[Debye Institute for Nanomaterials Science,Universiteit Utrecht]
{Soft Condensed Matter, Debye Institute for Nanomaterials Science,
 Universiteit Utrecht, Princetonplein 5, 3584 CC Utrecht, The Netherlands}
\author{Marijn A. van Huis}
\affiliation[Debye Institute for Nanomaterials Science,Universiteit Utrecht]
{Soft Condensed Matter, Debye Institute for Nanomaterials Science,
 Universiteit Utrecht, Princetonplein 5, 3584 CC Utrecht, The Netherlands}
\email{M.A.vanHuis@uu.nl}
\title{Strong long-range relaxations of structural defects in graphene simulated using a new semi-empirical potential}

\begin{document}

\begin{abstract}
We present a new semi-empirical potential for graphene, which includes also an out-of-plane energy term. This novel potential is developed from density functional theory (DFT) calculations for small numbers of atoms, and can be used for configurations with millions of atoms. Our simulations show that buckling caused by typical defects such as the Stone-Wales (SW) defect extends to hundreds of nanometers. Surprisingly, this long-range relaxation lowers the defect formation energy dramatically - by a factor of $2$ or $3$ - implying that previously published DFT-calculated defect formation energies suffer from large systematic errors. We also show the applicability of the novel potential to other long-range defects including line dislocations and grain boundaries, all of which exhibit pronounced out-of-plane relaxations. We show that the energy as a function of dislocation separation diverges logarithmically for flat graphene, but converges to a constant for free standing buckled graphene. A potential in which the atoms are attracted to the 2D plane restores the logarithmic behaviour of the energy. Future simulations employing this potential will elucidate the influence of the typical long-range buckling and rippling on the physical properties of graphene.
\end{abstract}

\section{INTRODUCTION}
Graphene has emerged as one of the most remarkable new materials of the last decade due to its extraordinary and unusual electronic, mechanical, optical and thermal properties.\cite{Novoselov2004,Novoselov2005,Balandin2008,Geim2009,Soldano2010,Lee2008} These properties make graphene an important component in the nanotechnology and semiconductor industry.\cite{Zhang2005,Geim2007,Yazyev2014} For these applications, however, defects need to be closely controlled since they  can significantly alter the physical and chemical properties of graphene.\cite{Mesaros2009,Chen2011} 

We are interested here in pronounced out-of-plane deformations (buckling) that arise naturally in the presence of defects in the two-dimensional (2D) structure of graphene.\cite{Fasolino2007,Katsnelson2008,Bhowmick2010,Zhang2014}  These deformations occur already in the presence of intrinsic defects, the simplest defects in graphene. We study zero-, one- and two-dimensional defects. Specifically, we consider a Stone-Wales (SW) defect, a dislocation line caused by the separation of two pairs of pentagon-heptagon rings, and grain boundaries.
For a review on intrinsic defects in graphene, we refer to the work of Meyer {\it {et al.}} \cite{mm} and the review by Banhart {\it {et al.}} \cite{Banhart2011}. 

An SW defect consists of two pentagons and two heptagons.\cite{Stone1986} Experimentally, SW defects are formed via rapid quenching from high temperature or under irradiation. \cite{mm} When four hexagons are converted into a pair of pentagons and heptagons in order to create an SW defect, the change in bond angles and bond lengths is significant, forcing buckling to decrease the induced strain. 

Buckling defects have already been studied in sheets and tubes. Georgii {\it {et al.}}, for example, have reported three different out-of-plane buckling modes and their activation barrier in carbon nanotubes (CNTs).\cite{Samsonidze2002} The formation and activation energies of a single SW defect have been calculated using various methods like quantum molecular dynamics calculations with an empirical potential,\cite{BuongiornoNardelli1998} DFT calculations, \cite{Li2005,Ma2009} and using a combination of semiempirical {\it {ab initio}} method and an amber force field.\cite{Zhou2003}

Since buckling depends on the defect rotation angle and shear force, defect orientation is important.\cite{Wang2013,Yazyev2010}. The most stable buckling mode for an SW defect in graphene  is the mode where one pentagon ring moves above the plane and the other below the plane at the dislocation core.\cite{Ma2009,Shirodkar2012} Moreover, and in agreement with the Mermin-Wagner theorem that states that the energy should diverge logarithmically and dislocations should attract each other in a two-dimensional sheet,\cite{Mermin1966,Mermin1968} Samsonidze {\it{et al.}} have shown that the attraction potential increases logarithmically as a function of the distance between dislocations.\cite{Samsonidze2002a}

In this paper, we further characterize this attraction potential in graphene's both flat and buckled modes. Experimental evidence shows that out-of-plane buckling in the graphene due to the dislocation dipole is quite significant and has a long-range effect that extends well beyond the size accessible to DFT calculations.\cite{Warner2012,Warner2013,Lehtinen2013} It is necessary, therefore, to turn to empirical potentials to study this effect.  While empirical potentials for carbon have already been reported in the literature, such as the Tersoff-Brenner \cite{Tersoff1988,Brenner1990} and the Stillinger-Weber potentials \cite{Stillinger1985}, they have not been fitted directly to graphene. Here, we turn to a simpler harmonic form based on the  Keating potential \cite{Keating1966} and Kirkwood potential \cite{Kirkwood1939} to describe the covalent bonds. We add an out-of-plane energy term, that is missing in all these empirical potentials, which allows us to better reproduce experimental and {\it{ab initio}} results, and to better understand the effects of intrinsic defects on the graphene sheet structure.

We first present the semi-empirical potential as fitted from DFT-calculations, and show its good performance in predicting structural properties of graphene, such as the elastic constants. Using this novel potential, we then study the effect of defects themselves on buckling. In particular, we look at a single SW defect and at the buckling caused by the spliting of an SW defect into a dislocation pair. We find that the elastic energy does not diverge logarithmically for buckled structures and we propose a mechanism to restore this logarithmic behavior. We further demonstrate the applicability of the potential by simulating grain boundaries in graphene for both flat and buckled modes.

\section{METHOD}
\subsection{Empirical potential}

Harmonic potentials such as Kirkwood's\cite{Kirkwood1939} and Keating's \cite{Keating1966},  represent a cheap and surprizingly accurate approach to study elastic deformation in fully connected covalent systems, even when fully disordered, with only a few fitted parameters.\cite{Wooten1985,Barkema2000} These simple potentials also allow us to better understand the origin of the various energy contributions. 

For semiconductors, these potentials typically include two energy terms ---a two-body bond-stretching and a three-body angular contribution--- and offer similar predictive quality, differing only in the details of implementation. While the Keating potential is more commonly used for computational convenience,   Kirkwood's representation has the advantage of providing a full separation between bond-stretching and angular contributions.\cite{Mousseau1992} Because, with current computing capabilities, cost difference between these representations is not an issue, we retain the second form, for formal convenience. 

For a two-dimensional hexagonal network, Kirkwood's potential is written as 
\begin{equation} \label{eq:Potential1}
E_0=\frac{3}{16}\frac{\alpha}{d^2}\sum_{i,j}(r^2_{ij}-d^2)^2+\frac{3}{8}\beta d^2 \sum_{j,i,k}(\theta_{j,i,k}-\frac{2 \pi}{3})^2 
\end{equation}
where $\alpha$ and $\beta$ are parameters fitted to the bulk and shear modulus, and the $2\pi/3$ term enforces 120 degree angles between the bonds.

To allow for deformation in the third dimension, here we introduce an additional term that imposes a harmonic restoring force in addition to the fourth-order energy correction that comes out from the bond-stretching pair term. The final potential is therefore written as 
\begin{equation} \label{eq:Potential}
E=\frac{3}{16}\frac{\alpha}{d^2}\sum_{i,j}(r^2_{ij}-d^2)^2+\frac{3}{8}\beta d^2 \sum_{j,i,k}(\theta_{j,i,k}-\frac{2 \pi}{3})^2 + \gamma \sum_{i,jkl}r^2_{i,jkl}
\end{equation}
with $d = 1.420$ \AA, the ideal bond length for graphene, and the other parameters being extracted from DFT calculations are: $\alpha = 26.060$~eV/\AA$^{2}$, $\beta = 5.511$~eV/\AA$^{2}$  and $\gamma = 0.517$~eV/\AA$^{2}$. Here, $r_{i,jkl}$ is the distance between atom $i$ and the plane through the three atoms $j$, $k$ and $l$ connected to atom $i$. 

\subsection{Obtaining the fitting parameters from DFT calculations}

Parameters for the empirical potential are fitted from DFT calculations using the the first-principles Vienna {\it{Ab initio}} Simulation Package (VASP) \cite{Kresse1993,Kresse1996}. To describe buckling accurately, we used a van der Waals functional \cite{KlimeAa2009}, which is shown to work well for solids \cite{Klimefmmodeheckslsesi2011}. The van der Waals functional was formulated by Dion and co-workers\cite{Dion2004}. The cut-off energy of the wave functions was $400$ eV. The cut-off energy of the augmentation functions was about $650$ eV. The electronic wave functions were sampled on a $6\times8\times1$ grid with $24$ k-points in the irreducible Brillouin zone (BZ) of graphene using the Monkhorst and Pack method \cite{Monkhorst1976}. Structural optimisations were performed for both lattice parameters and coordinates of atoms. Different k-meshes were tested for a primitive graphene cell, and the cut-off energies for the wave functions and augmentation wave functions were also tested in order to ensure good energy convergence ($< 1$ meV/atom).

The various parameters were fitted by imposing elastic deformation onto a graphene sheet. The value of $\alpha$ was obtained by fitting the quadratic energy evolution as a function of uniform deformations on a 50-atom monoclinic perfectly crystalline graphene sample as shown in Figure 1(a). Since this is a homogeneous compression, with deformation constrained to the plane, only the two-body term in the potential contributes to the energy. The obtained value,  $\alpha = 26.060$~eV/\AA$^{2}$, is in good agreement with the reported  experimental value of $25.880$~eV/\AA$^{2}$.\cite{Mohr2007,Kumar2012}

\begin{figure}  
\includegraphics[width=0.9\textwidth]{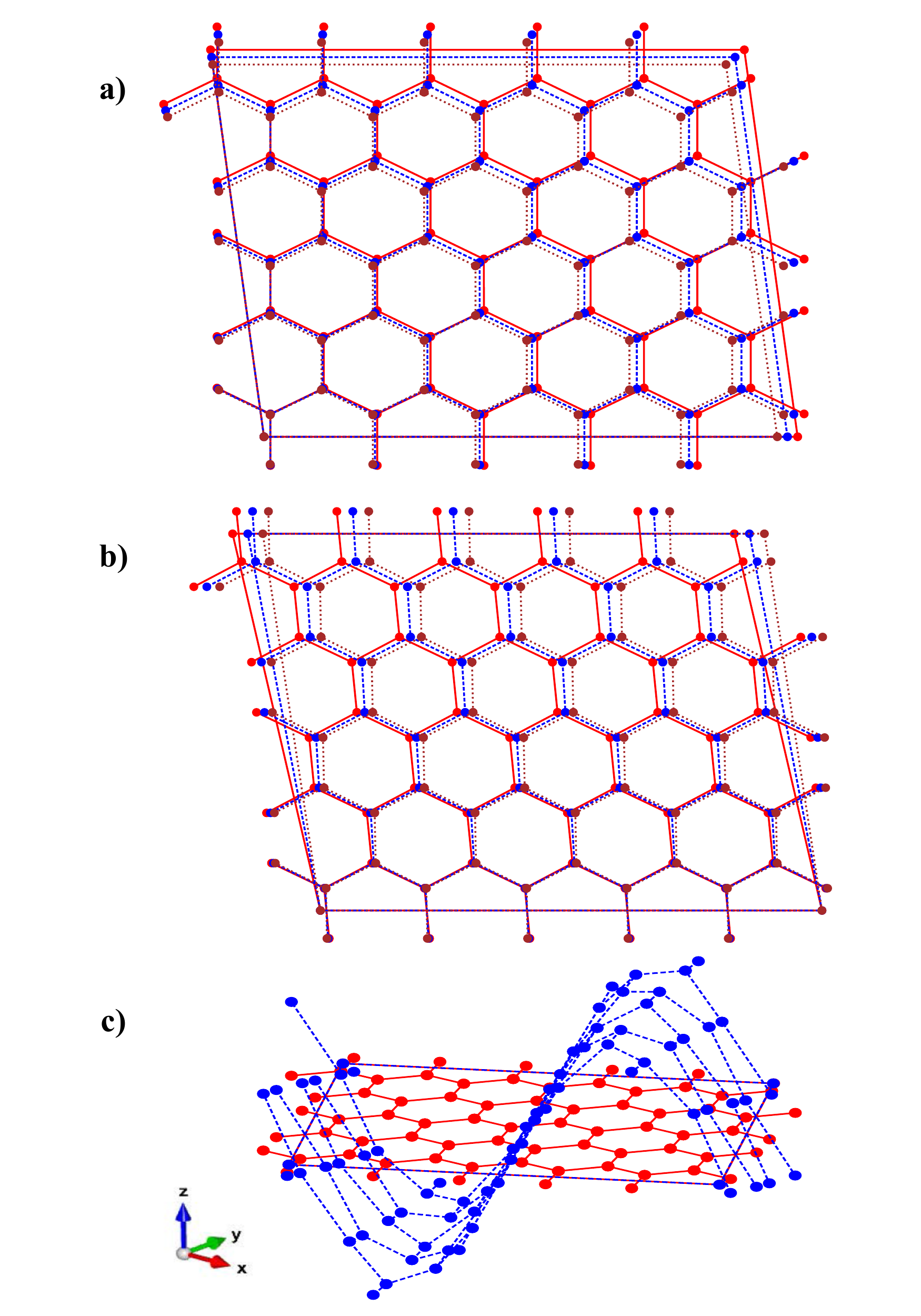}
\caption{50-atom samples in which different kinds of elastic modes are excited. (a) Homogeneous compression of the box. (b) Shearing of the box at constant area. (c) Sinusoidal displacement in $z$-direction. These excitations are used to determine the parameters $\alpha$, $\beta$ and $\gamma$ in our potential Equation 2.}
\label{fig:Figure_1}
\end{figure}

To obtain $\beta$, we repeat the procedure, this time by shearing the box, i.e. changing the angle between the periodicity vectors $L_{x}$ and $L_{y}$. During the process of shearing the box, the area is kept constant by scaling the box length in y direction accordingly, as shown in Figure 1(b). The energy as a function of shearing is also well fitted by a quadratic equation, leading to $\beta = 5.511$~eV/\AA$^{2}$.

Finally, $\gamma$ is obtained through sinusoidal displacements in $z$-direction given as \begin{equation} \label{eq:sine}
z_i = A \sin \left(\frac{2\pi k x_i}{L_{x}}\right).
\end{equation}
The amplitude $A$ of the displacement is varied from -0.1 to 0.1 \AA~as shown in Figure 1(c). Deformed samples are allowed to relax laterally at the fitted value of $\alpha$ and $\beta$.

The energy of these laterally relaxed samples is then fitted to
\begin{equation} \label{eq:gamma}
E=E_0 N +\gamma^l A^2 \left(\frac{k}{L_{x}}\right)^2 N.
\end{equation}
Here, N is the total number of the atoms in the sample and $\gamma^l$ is a fitting parameter. The energies of these samples were also computed by DFT, employing the same boundary conditions and fitted to Equation 4. Whereas the value thus found, $\gamma = 0.517$~eV/\AA$^{2}$, is almost 2 orders of magnitude smaller than $\alpha$ and one order of magnitude smaller than $\beta$, it remains significant, demonstrating the need to include such a harmonic term in the potential energy. 

To demonstrate the applicability of this novel potential, we study the deformation caused by intrinsic defects in graphene. The effect of defects on the geometry of a graphene sheet is examined in three parts. First, we look at the long-range deformation associated with the creation of an SW defect, representing a dislocation dipole. We then consider two types of line defects; split dislocation pairs and grain boundaries.

\section{RESULTS AND DISCUSSION}
\subsection{The Stone-Wales defect}

\begin{figure}  
\includegraphics[width=1.0\textwidth ]{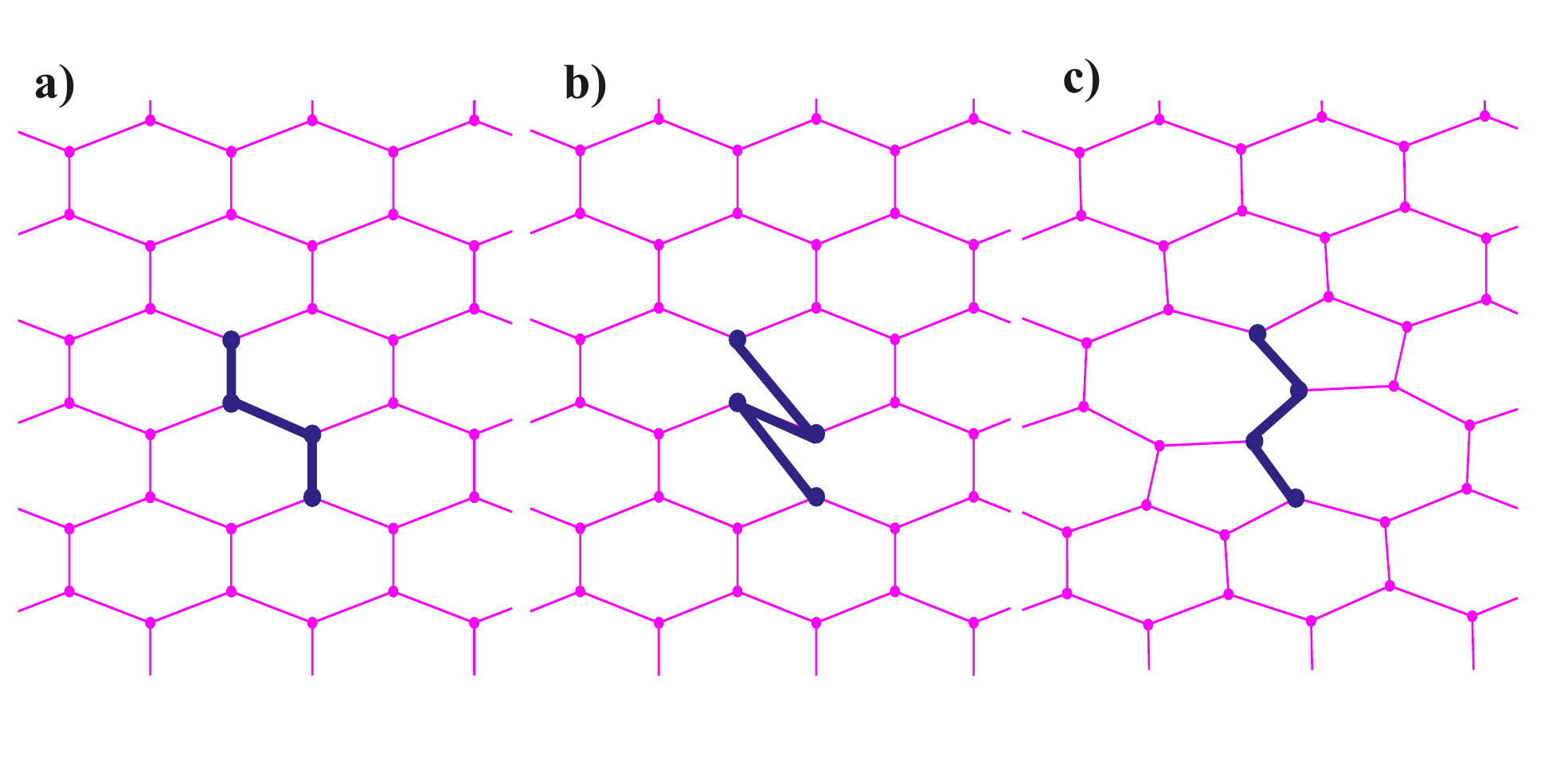} 
\caption{Creation of single Stones-Wales defect by bond transposition. (a) A perfect hexagonal arrangement. (b) Bonds are rearranged to create the topology of a defect. (c) The atomic positions are relaxed to generate a single SW defect.}
\label{fig:Figure_2}
\end{figure}

A single SW defect (5-7-7-5 member rings) \cite{Stone1986}  is created in a perfectly hexagonal lattice via a bond-switching move as shown in Figure \ref{fig:Figure_2}. After the bond transposition, the sample is relaxed with the novel potential, both when constraining the atoms in the 2D plane and when allowing full 3D relaxation. Several samples containing different numbers of atoms are generated to study the formation energy of SW defect in both flat and buckled modes. 

We first compare the SW-defect formation energy for the DFT and the semi-empirical
description using a small 42-atom model with a fixed area, to maintain a constant electronic density. Periodic boundary conditions apply in all cases. For the 2D relaxation, we find a good agreement between formation energy of 7.95 eV as obtained from DFT calculations and the value of 7.50 eV obtained when using the Kirkwood-like potential (Table \ref{Table 1}). To trigger relaxation in the third dimension, the symmetry is broken by the addition of a small out-of-plane displacement ($0.01$\AA) with alternating sign to the atoms participating to the SW defect. This buckling initialization was adopted to avoid metastable configurations. This leads to a staggered configuration that is of lower energy with atoms moving above and below the plane during the relaxation.\cite{Samsonidze2002}. As seen in Table \ref{Table 1}, the additional degree of freedom in the out-of-plane direction reduced the strain by more than 2 eV in the small box.

\begin{center} 
\begin{table}
\caption{Formation energy of an SW defect calculated using DFT and using our semi-empirical potential. These values are obtained for a 42-atom sample at fixed crystalline density. The value $E_{2d}$ is obtained in a planar geometry, and the value $E_{3d}$ is obtained with out-of-plane relaxation. $\Delta E$ is the energy difference between these two configurations after relaxation.}
\label{Table 1}
\begin{tabular}{c c c c}
\hline
Method & $E_{2d}$ (eV) & $E_{3d}$ (eV) & $\Delta E$ (eV) = $E_{2d}$ - $E_{3d}$ \\ \hline
DFT & 7.952 & 5.933 & 2.019  \\ 
Semi-empirical potential & 7.496 & 5.274 & 2.222 \\ \hline
\end{tabular}
\end{table} 
\end{center}

We also study the energy convergence of the SW defect with system size, looking at systems ranging from $N=42$ to 69200 atoms. This time, we allow the box size to relax, which also decreases the elastic energy by more than 2 eV (Table  \ref{Table 2}). Nevertheless, size convergence for both relaxation types is slow and, for 2D relaxation, non-uniform. In this case, the elastic energy associated with the SW defect is 5.96 eV for a 42-atom structure, and increases by 0.9 eV to 6.87 eV for a 680-atom system, and converges to 6.73 eV for a system containing 6696 atoms.  Overall, the formation energy for the 2D SW defect is in very good agreement with values reported in the literature calculated by DFT \cite{Ma2009} and using a combination of semiempirical {\it{ab-initio}} method and an amber force field \cite{Zhou2003} for small samples. 

\begin{center} 
\begin{table}
\caption{Formation energy of the SW defect calculated using our semi-empirical potential.  The value $E_{2d}$ is obtained in a planar geometry, and the value $E_{3d}$ is obtained with out-of-plane relaxation. In contrast to Table 1, the density is not fixed to its crystalline value but the box is allowed to relax (box lengths $L_x$ and $L_y$ as well as the angle between them). $\Delta E$ represents the energy difference between the two energy-minimized structures and $\Delta z_{max}$, the buckling height, defined as the difference between maximum and minimum $z$-direction co-ordinates.}
\label{Table 2}
\begin{tabular}{c c c c c}
\hline
Sample size(Number of the atoms) & $E_{2d}$ (eV) & $E_{3d}$ (eV) & $\Delta E$ (eV) = $E_{2d}$ - $E_{3d}$ & $\Delta z_{max}$ (\AA) \\ \hline
42 & 5.963 & 2.636 & 3.327 & 2.460 \\ 
336 & 6.642 & 2.814 & 3.823 & 3.226 \\ 
680 & 6.868 & 2.831 & 4.037 & 3.640 \\ 
1008 & 6.700 & 2.840 & 3.860 & 3.939 \\ 
1344 & 6.707 & 2.843 & 3.864 & 4.190 \\ 
3344 & 6.721 & 2.852 & 3.869 & 5.215 \\ 
6696 & 6.725 & 2.856 & 3.869 & 6.190 \\ 
17200 & 6.728 & 2.861 & 3.867 & 7.790 \\ 
34160 & 6.729 & 2.866 & 3.863 & 9.178 \\ 
69200 & 6.729 & 2.868 & 3.861 & 10.855 \\ \hline
\end{tabular}
\end{table} 
\end{center}

\begin{figure}  
\includegraphics[width=1.0\textwidth]{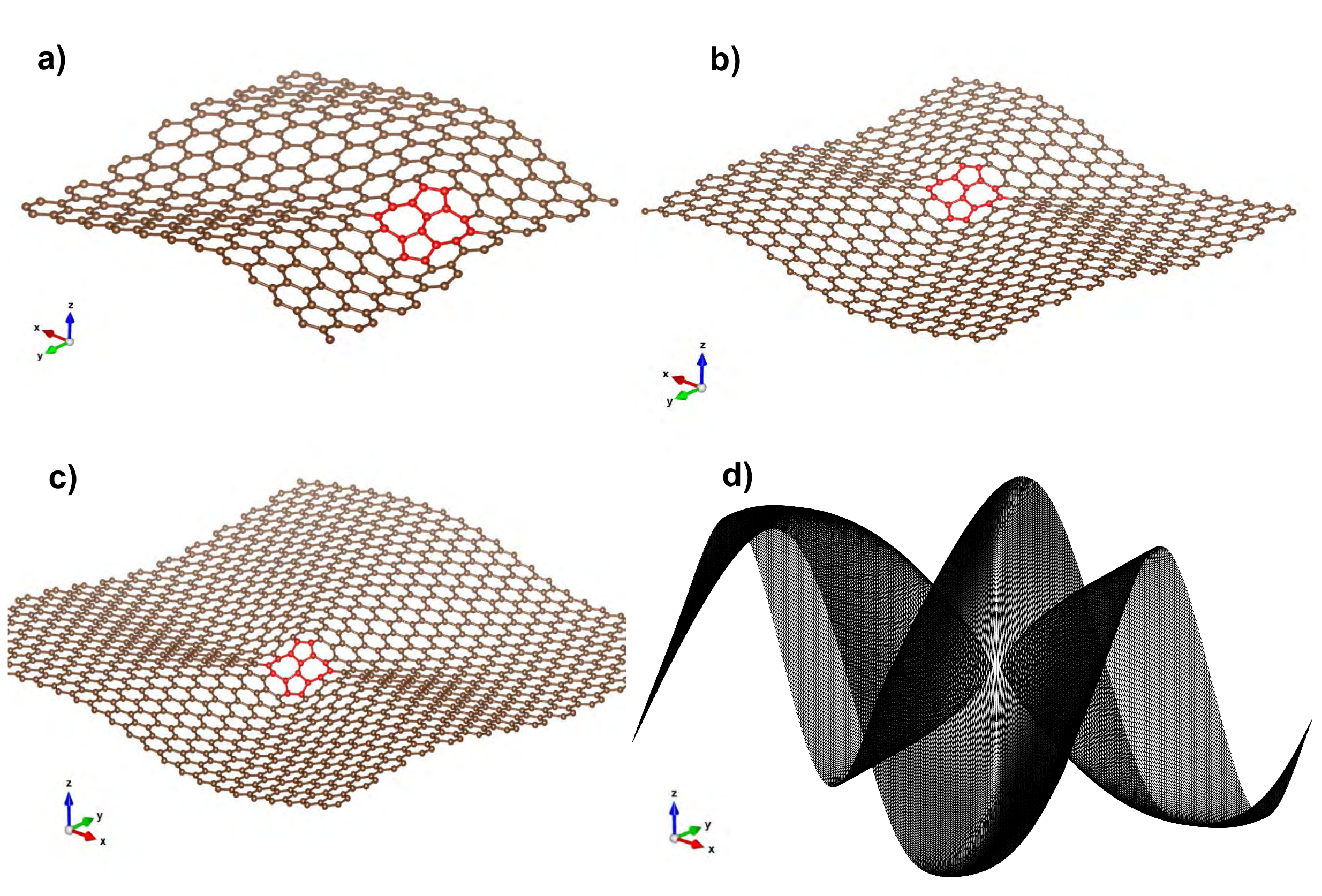} 
\caption{Well-relaxed buckled graphene samples having a single SW defect. Periodic boundary conditions apply. During the minimization process one pentagon ring goes above the plane and another goes below the plane. These out-of-plane displacements extend throughout the sample causing long-ranged buckling. (a) Sample with 336 atoms with $x$ and $y$ dimensions of $\pm$ 20 \AA~and bulging in $z$-direction up to $\pm$ 2 \AA. (b) Sample with 680 atoms with $x$ and $y$ dimensions of $\pm$ 25 \AA~and bulging in $z$-direction up to $\pm$ 2 \AA. (c) Sample with 1344 atoms with $x$ and $y$ dimensions of $\pm$ 30 \AA~and bulging in $z$-direction up to $\pm$ 2.5 \AA. (d) sample with 69200 atoms with $x$ and $y$ dimensions of $\pm$ 250 \AA~and bulging in $z$-direction up to $\pm$ 6 \AA.}
\label{fig:Figure_3}
\end{figure}

The energy ($E_{3d}$) for the buckled samples is significantly lower (around $3.86$ eV) than for the 2D-constrained models. The resulting 3D relaxation leads to a local minimum-energy structure for the SW defect, with one pentagon ring moving above the plane while the other moves below the plane.  The bond connecting both rings has the tendency to be at an angle of inclination of approximatly $18^0$ with respect to the plane. This  relaxation is accompanied by significant buckling --- varying from 2.5~\AA\, for the smallest box to almost 11 \AA\, for the largest system --- that continues to grow with system size, a phenomenon that cannot be captured with small DFT calculations (see Figure  \ref{fig:Figure_3}). Note that the formation energy of the 3D SW defect for the largest system is 2.87 eV, which is more than two times lower than the DFT-calculated value of 5.93 eV for the small 42-atom sample. This clearly shows that long-range relaxation can drastically reduce defect formation energies, and that these effects cannot be studied using only DFT simulations. 

\begin{figure}   
\centering
\includegraphics[width=1.0\textwidth]{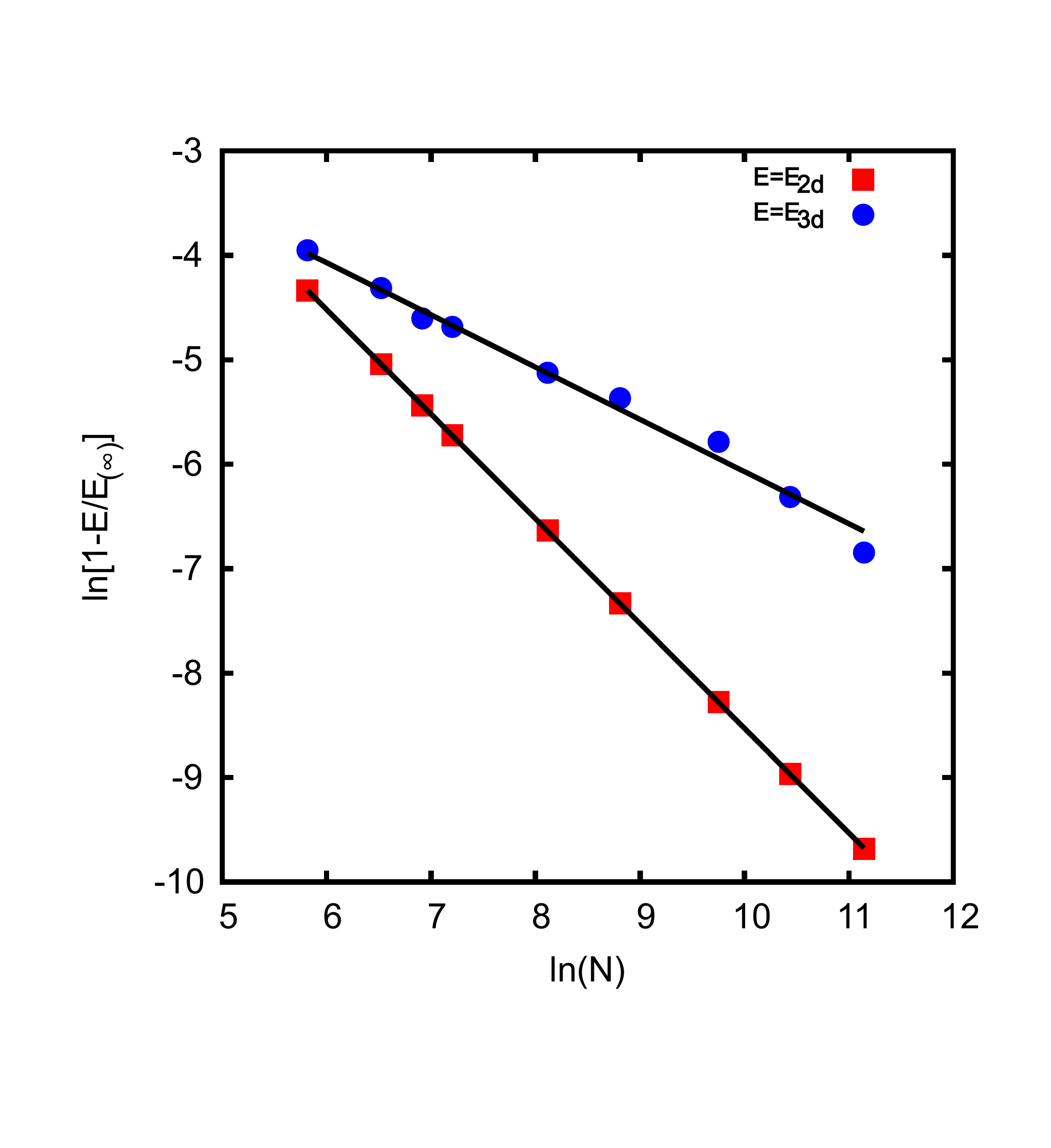}             
\caption{Effect of finite box size on SW formation energy, with and without buckling. We expect the energy to scale as $E(N)=E(\infty) - C N^{-p}$. Hence, $\ln[1-E(N)/E(\infty)] = \ln(C/E(\infty)) - p \ln(N)$. The data fall on straight lines with our extrapolated energies $E_{2d(\infty)}$=6.729 eV and $E_{3d(\infty)}$=2.869 eV, with slopes $-1.0$ and $-0.5$, respectively. This suggests that finite-size corrections are $\sim 1/L^2$ and $\sim 1/L$, respectively.}
\label{fig:Figure_4}
\end{figure}

Figure \ref{fig:Figure_4} shows $\ln\left[1-\frac{E}{E_{(\infty)}}\right]$ as a function of $\ln(N)$ with $E_{2d(\infty)}$ and $E_{3d(\infty)}$ determined by extrapolation to be 6.729 eV and 2.869 eV, respectively. These data points were fitted by a straight line. In the 2D case the straight line has a slope of $-1.0$ and intercept of $1.48$. In the 3D case, we observe rather a slope of $-0.5$ and an intercept of $-1.07$. Since $\sqrt{N}$ is proportional to the sample size ($L$) we can conclude that $E_{3d}$ has a finite size correction that scales as $1/L$ in the energy, whereas $E_{2d}$ has a finite size correction of $1/L^2$ in the energy. 

\subsection{Line defect: dislocation formed by a separated Stone-Wales defect}

\begin{figure}
\includegraphics[width=1.0\textwidth]{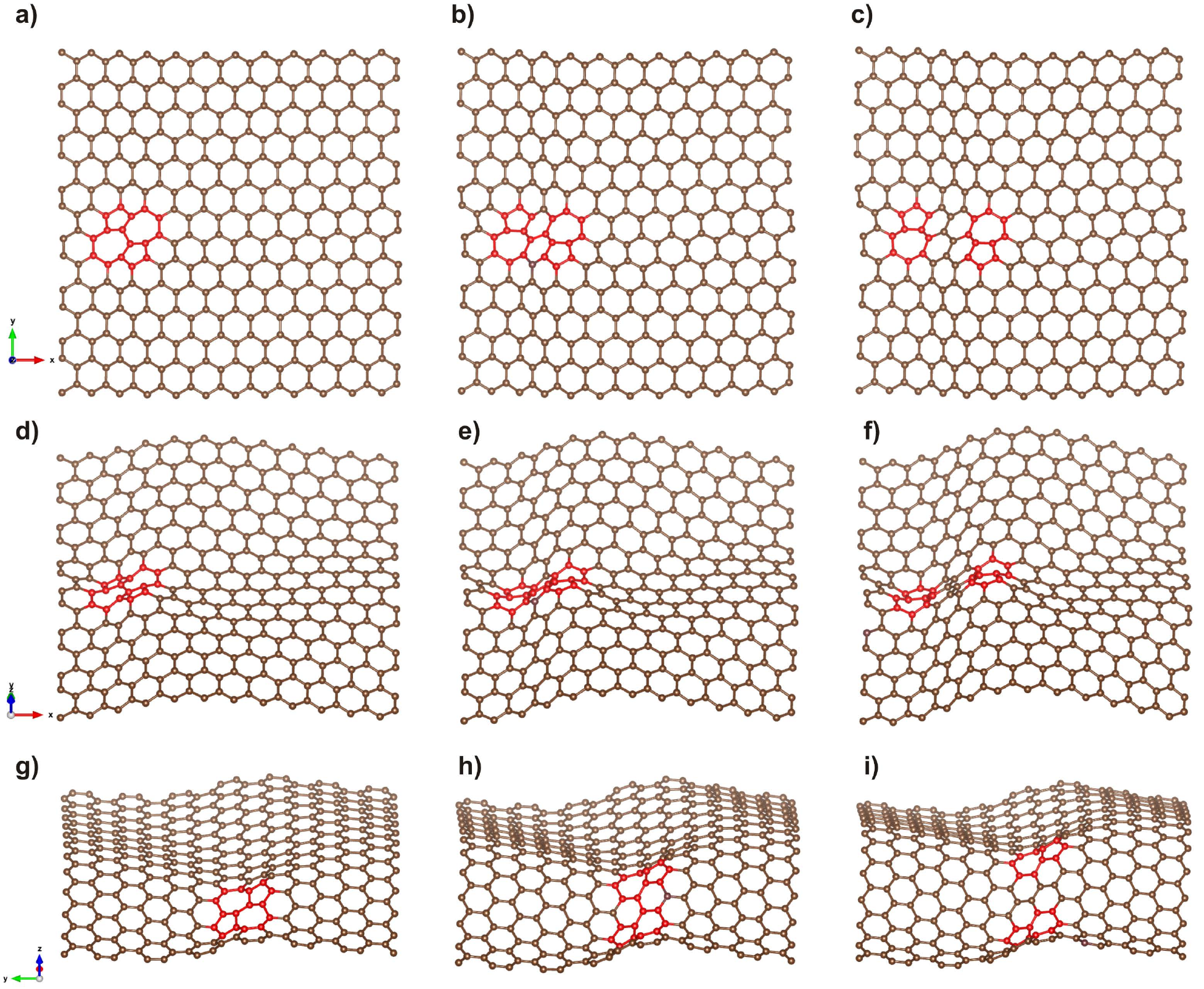}
\caption{Structural changes induced by a (5-7) dislocation after 3D relaxation. (a,d,g) Samples with a single Stone-Wales defect at various viewing angles (b,e,h) Dislocations separated by one hexagonal ring ($\Delta=1$) (c,f,i) Dislocations seperated by 2 hexagonal rings ($\Delta=2$).}
\label{fig:Figure_5}
\end{figure}

An SW defect represents a dislocation dipole. The dislocation pairs (pentagon-heptagon rings) can then be separated by inserting hexagon rings in between them, leading to the creation of a line defect as shown in Figure \ref{fig:Figure_5}.

\begin{figure} 
\includegraphics[width=0.95\textwidth]{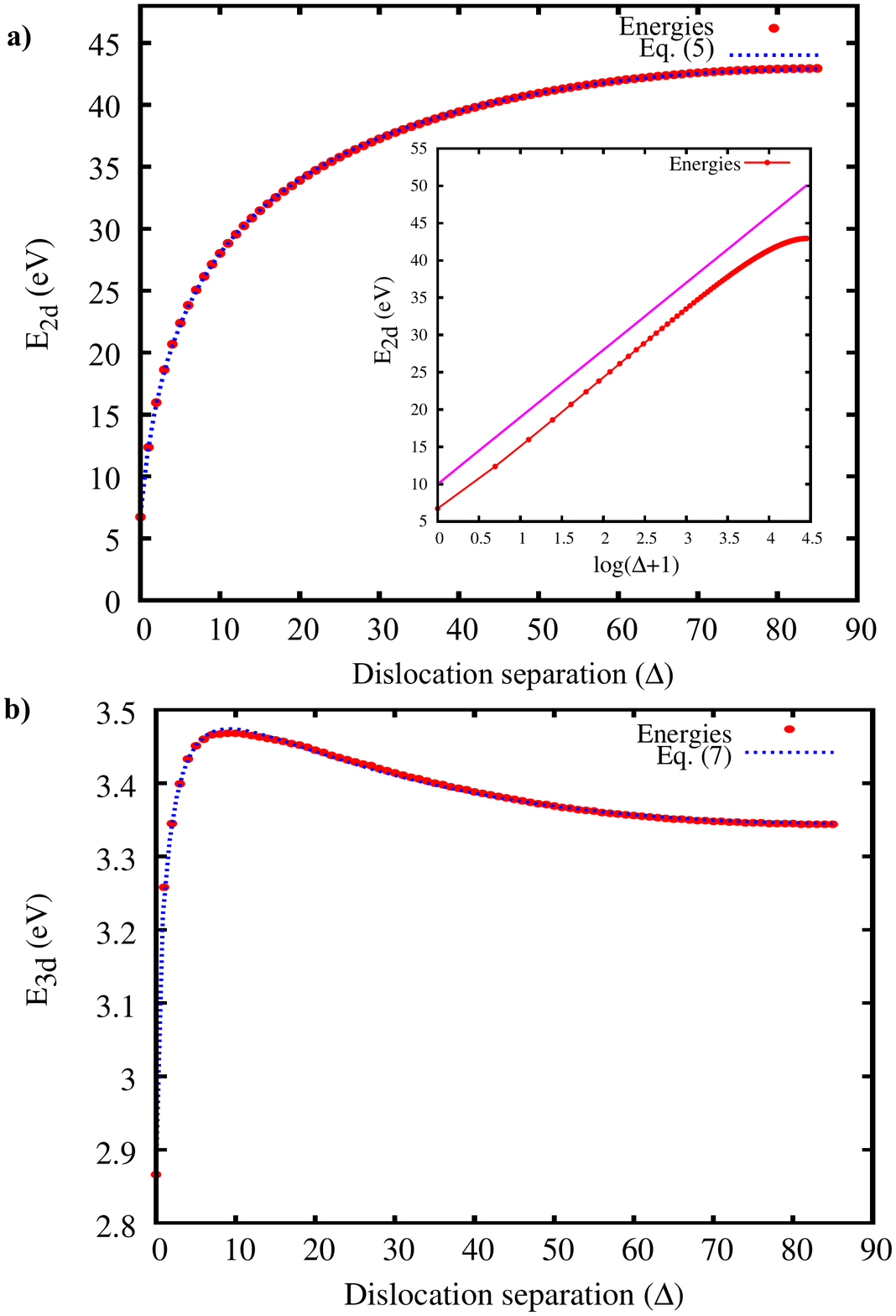}
\caption{Energy ($E(\Delta)$) as a function of dislocation separation ($\Delta$) in a 69200-atoms sample. (a) Without out-of-plane relaxation, the data points are well fitted by Equation 5 (line).
The inset shows the logarithmic increase of the energy with $\Delta$, which flattens once $\Delta$ approaches half the box size; the straight line is drawn to guide the eye. (b) Relationship found when out-of-plane relaxation is allowed, this time fitted by Equation 7 (line).} 
\label{fig:Figure_6}
\end{figure}

The energy as a function of dislocation separation $\Delta$ is shown in Figure \ref{fig:Figure_6}. In the 2D case where atomic motion is constrained to the $xy$-plane, the energy  diverges logarithmically as a function of $\Delta$; this is consistent with the Mermin-Wagner theorem \cite{Mermin1966,Mermin1968} and the KTHNY Theory \cite{Kosterlitz1973,Nelson1979,Young1979} (see Figure 6(a)).
This energy evolution was fitted to the equation
\begin{equation} \label{eq:logarithmic}
E_{2d}(\Delta)=a + b \ln\left[\frac{1}{\Delta+c}+\frac{1}{P-\Delta+c}\right]^{-1}
\end{equation}
where $a$, $b$ and $c$ are the fitting parameters and $P$ is the total number of the hexagonal rings in the direction of $\Delta$. For the 69200-atom sample, we consider P=171 and $\Delta_{max}$=85.

As the symmetry is broken through the introduction of small perturbations along the $z$-axis, relaxation is more effective and the energy as function of dislocation separation converges to a constant with corrections scaling as $\frac{1}{\sqrt{\Delta}}$ for large values of $\Delta$. Lehtinen {\it{et al.}} have shown experimentally that the evolution of dislocations in 2D systems is governed by long-range out-of-plane buckling.\citep{Lehtinen2013} Our simulations also indicate that line defects in graphene (a dislocation pair) can have significant impact on buckling. 

For our $69200$-atom sample, for example, buckling increases dramatically with respect to dislocation separation. The buckling height increases from 10.9 \AA~for $\Delta=0$ to 46.5 \AA~for $\Delta=85$. Moreover, the interaction between dislocations, in this case, is not monotonic: whereas in 2D interaction remains attractive at all distances, the interaction changes sign when 3D-relaxation is allowed, and becomes repulsive after some threshold distance as can be clearly seen in Figure 6(b). These effects could not be observed numerically before due to the small sample sizes used previously. The behavior of the elastic energy as a function of distance between two dislocations when 3D-relaxation is allowed can be described by
\begin{equation}
f(\Delta)= a+\frac{c+b\sqrt{\Delta}}{d+\Delta}
\end{equation}
and 
\begin{equation} \label{eq:3d}
E_{3d}(\Delta)= f(\Delta)+f(P-\Delta)
\end{equation}
where $a$, $b$, $c$ and $d$ are the fitting parameters and $P$ is the system size defined as total number of the six member rings in the direction of $\Delta$. For large values of $\Delta$, the force acting between the dislocation pairs can be written together for 2D and 3D cases as
\begin{equation} \label{eq:force}
F(\Delta) \propto \frac{1}{\Delta^{D/2}}
\end{equation} 
with $D$ the dimensionality of the system.

\begin{figure} 
\includegraphics[width=1.0\textwidth]{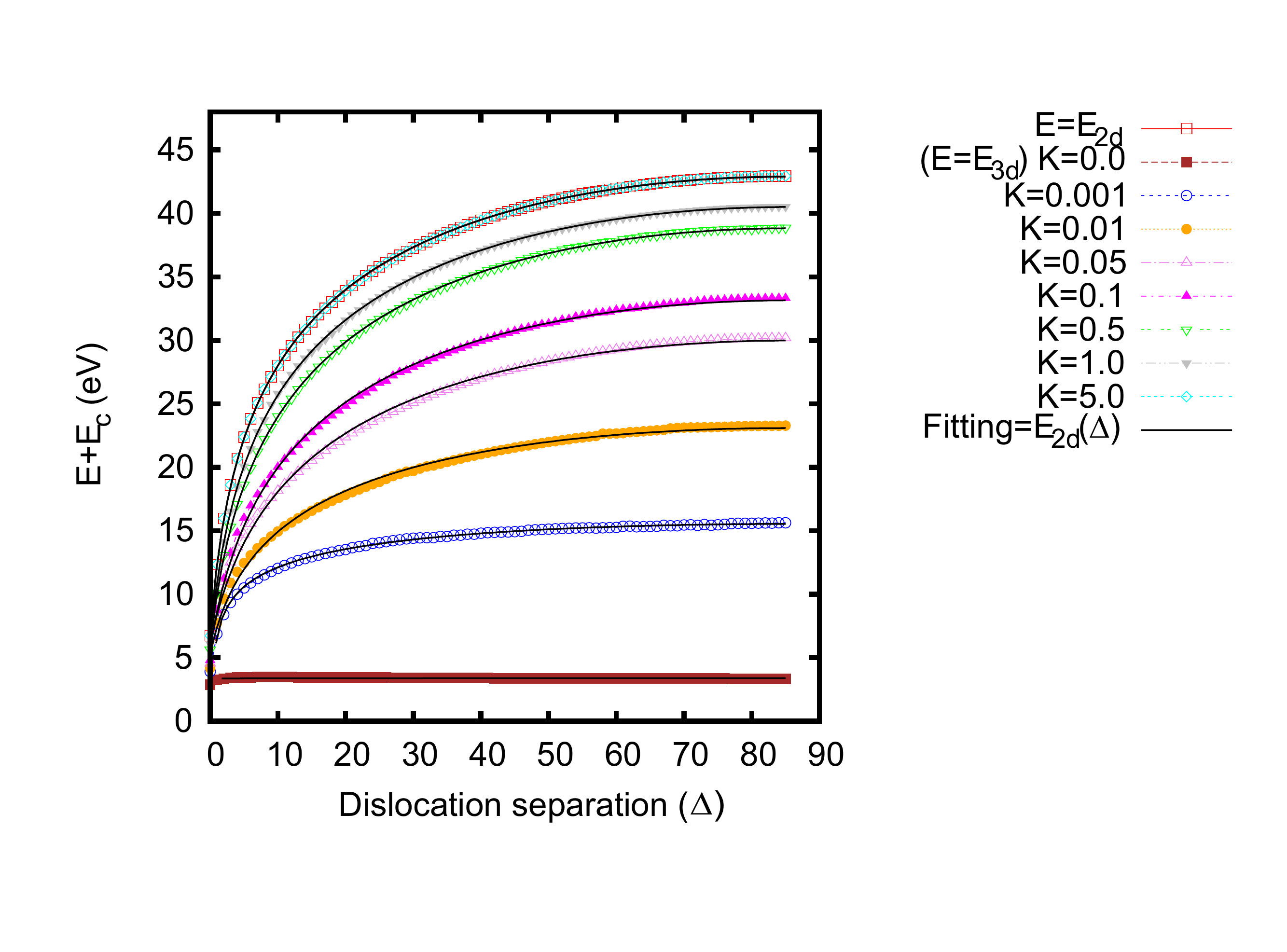}
\caption{Energy ($E(\Delta)$) as a function of dislocation separation ($\Delta$) in a 69200-atoms sample for different values of strengths $K$ (eV \AA$^{-2}$) of the confining potential. Note that $K=0$ (see Equation 9) is for free standing graphene, while $K\rightarrow \infty$ is the limit to fully planar graphene. This figure shows that at any nonzero $K$, the energy diverges logarithmically, whereas for $K=0$ (free standing graphene), the energy converges to a constant.} 
\label{fig:Figure_7}
\end{figure}

\begin{figure}  
\includegraphics[ width=1.0\textwidth]{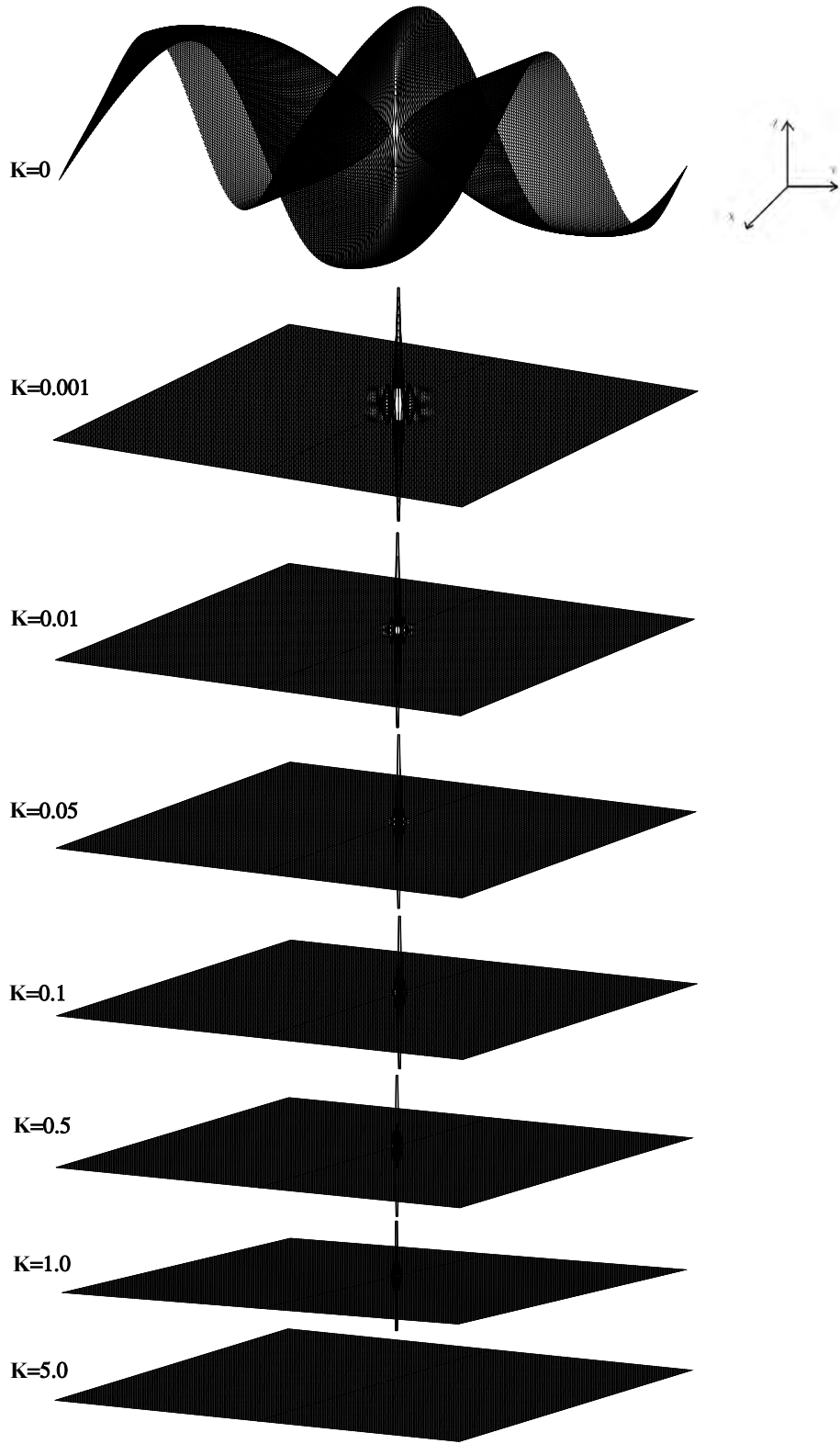} 
\caption{Out-of-plane structure of an SW defect, as a function of the strength $K$ (eV \AA$^{-2}$) of the confining potential for a sample with 69200 atoms. Note that even a very small value of $K$ already localises the out-of-plane effect. Values of $z$-direction displacements are given in Table 3.}
\label{fig:Figure_8}
\end{figure}

To simulate the effect of substrate on the buckling of graphene, we add a harmonic confining energy term in our potential. The logarithmic divergence of the energy as a function of dislocation separation is again restored by this harmonic confinement energy term that is given as:
\begin{equation} \label{eq:spring}
E_{c}= K \sum_{i=1}^N z^2_i.
\end{equation}
$K$ is the pre-factor for this harmonic term (eV \AA$^{-2}$), $N$ is number of the atoms present in the sample and $z_i$ is the normal-to-plane co-ordinate of the atom. So, all the buckeled graphene samples relaxed back to the 2D plane with the confinement potential given in Equation 9 for different values of $K$. The energy as a function of $\Delta$ was calculated and plotted as shown in Figure \ref{fig:Figure_7}. It is evident from the plot that for any non-zero value of $K$ the logarithmic behaviour of the energy again restores and at $K=5$ eV \AA$^{-2}$ the energy plot overlaps with the $E=E_{2d}$ plot. With this additional confining term $E_c$ in the potential, samples again have the tendency to become flat with some small $z$-direction fluctuations at the core of the defect. It has been observed that for the case $K=0$ eV \AA$^{-2}$ (free standing buckled graphene), the $z$-direction displacement increases with increasing dislocation separation (Table \ref{Table 3}). However when a non-zero harmonic confinement potential is applied, $z$-direction displacements become very small and also very local near the dislocation core (Figure \ref{fig:Figure_8}). These buckling heights become smaller and smaller with strong harmonic confinement. Therefore, logarithmic behavior of the energy as a function of dislocation separation restores again with harmonic confinement. 

The $K$ value determines the lateral localization of the buckling. We have analyzed the lateral extension of the buckling at different values of $K$. At $K=0.001$ eV \AA$^{-2}$, the lateral extention of buckling is around 40 \AA, whereas this decreases to 20 \AA ~ for $K=0.01$ eV \AA$^{-2}$ and 12 \AA ~ for $K=0.05$ eV \AA$^{-2}$. Tison {\it{et al.}} \cite{Tison2014} have shown the buckling arising due to the grain boundaries extends to typically 5 to 20 \AA ~ on SiC substrate. So, we estimate that a value of K=0.001 to 0.01 eV \AA$^{-2}$ is a reasonable estimate to simulate the presence of a substrate. 

\begin{center} 
\begin{table}
\caption{Values of buckling height $\Delta z_{max}$ in \AA ~(difference between maximum and minimum $z$-direction co-ordinates) for different values of $K$ (eV \AA$^{-2}$) (see Equation 9) at different dislocation separation distances ($\Delta$) for a sample with 69200 atoms.}
\label{Table 3}
\begin{tabular}{c c c c c c c c c}
\hline
$\Delta$ & $K=0$ & $K=0.001$ & $K=0.01$ & $K=0.05$ & $K=0.1$ & $K=0.5$ & $K=1.0$ & $K=5.0$ \\ \hline
0  & 10.855 & 2.461 & 2.138 & 1.751 & 1.580 & 0.975 & 0.705 & 0.000 \\ 
2  & 17.481 & 4.361 & 3.561 & 2.597 & 1.722 & 1.237 & 0.987 & 0.000 \\ 
5  & 21.380 & 4.444 & 3.459 & 2.757 & 2.437 & 1.309 & 1.048 & 0.089 \\ 
10 & 24.463 & 4.273 & 3.485 & 2.828 & 2.521 & 1.346 & 1.066 & 0.113 \\ 
20 & 27.579 & 4.264 & 3.557 & 2.885 & 2.568 & 1.360 & 1.072 & 0.118 \\ 
30 & 29.140 & 4.453 & 3.584 & 2.921 & 2.575 & 1.365 & 1.073 & 0.120 \\ 
40 & 33.510 & 4.400 & 3.600 & 2.935 & 2.590 & 1.362 & 1.073 & 0.122 \\ 
50 & 38.300 & 4.393 & 3.610 & 2.941 & 2.632 & 1.379 & 1.076 & 0.123 \\ 
60 & 42.290 & 4.376 & 3.617 & 2.939 & 2.597 & 1.364 & 1.076 & 0.124 \\ 
70 & 44.973 & 4.388 & 3.593 & 2.942 & 2.594 & 1.383 & 1.076 & 0.124 \\ 
80 & 46.304 & 4.389 & 3.600 & 2.943 & 2.599 & 1.371 & 1.076 & 0.124 \\ 
85 & 46.488 & 4.450 & 3.626 & 2.942 & 2.599 & 1.363 & 1.076 & 0.124 \\ \hline
\end{tabular}
\end{table} 
\end{center}

\subsection{Grain boundaries (domains) in graphene}

To demonstrate the broad applicability of our newly developed potential, we also simulate grain boundaries in graphene. To generate a sample, we start with a completely random 2D sample (all sites are 3-fold connected) and allow it to evolve doing WWW bond transpositions \cite{Wooten1985} with respect to time at room temperature. 

The initial connectivity is generated as follows. We first generate a Voronoi diagram from the initial random configuration\cite{Voronoi1908}, whereby the boundaries between each voronoi point are defined by the crossing normals of the mid-point connected each voronoi point to their nearest neighbours. This allows us to generate an unbiased isotropic three-fold connected random network.

To generate a Voronoi network we start with $N/2$ random points in a square box and placed $8$ copies of this box around it to implement periodic boundary conditions. Then Voronoi vertices were identified and connented in order to make bonds in between them. Hence, a random sample having $N$ atoms is generated. 

Next, we use the improved bond-switching WWW-algorithm to evolve the system \cite{Barkema2000}, using our empiricial potential to describe its energy. After each bond-switch the system is fully relaxed and the configuration is accepted with a Metropolis probability given by 
\begin{equation}
P=\min\left[1,\exp\left(\frac{E_b-E_f}{k_BT}\right)\right].
\end{equation}
Here $E_{b}$ is the energy before the bond transposition and $E_{f}$ is the energy after the bond transposition.

\begin{figure}  
\includegraphics[ width=1.0\textwidth ]{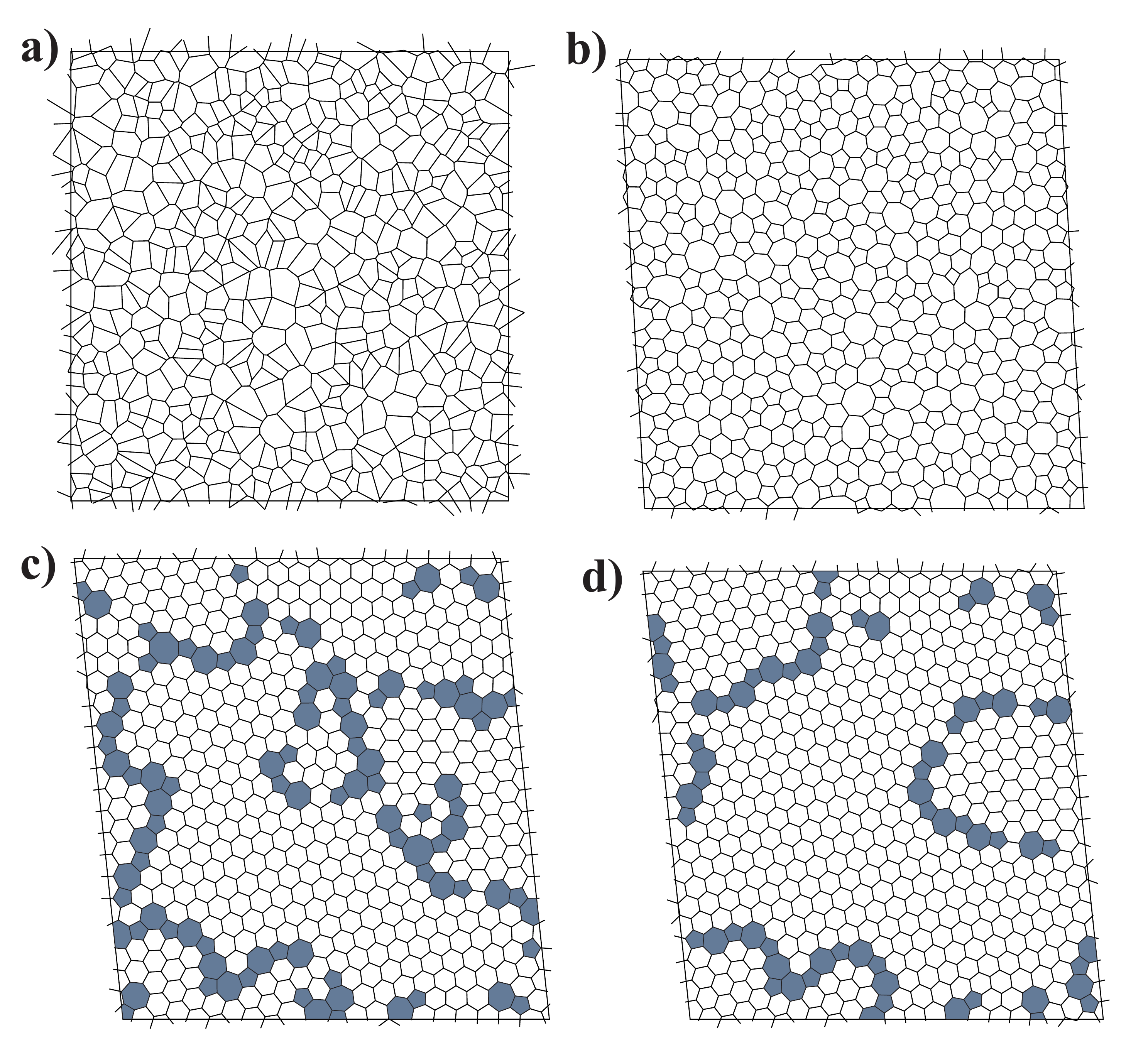} 
\caption{Structural evolution of a 1000-atom sample by bond transposition moves, confined to 
strictly planar configurations. (a) The starting sample: a random periodic Voronoi network which is unrelaxed. (b) Sample after minimization, and some minor manipulation to remove structural anomalies (t=0). (c) Sample having early evolved domains (t=25). (d) Sample having mature domains (t=100). The unit of time is $N$ bond transposition moves.}
\label{fig:Figure_9}
\end{figure}

\begin{figure}  
\includegraphics[width=1.0\textwidth ]{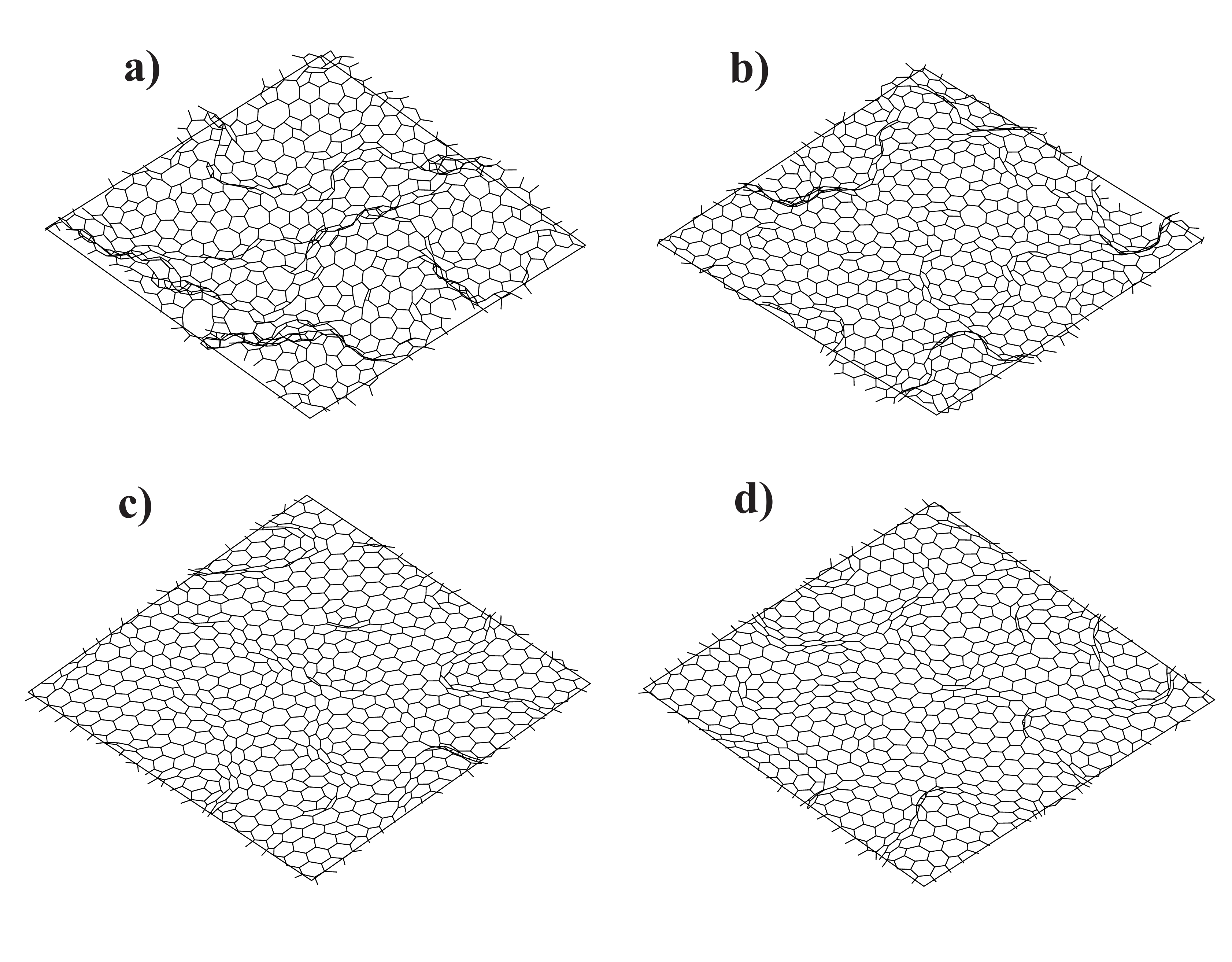} 
\caption{Structural evolution of a 1000-atom sample by bond transposition moves, with out-of-plane relaxation. (a) Starting sample, equal to that in figure 9(b), but with random out-of-plane displacements (after which it is relaxed) (t=0). (b) Sample having early evolved domains (t=5). 
(c) Sample having more mature domains (t=15). (d) Sample with mature domains (t=100). 
Here, $x$ and $y$ dimensions are 50 $\times$ 50 \AA$^2$~with bulging in $z$-direction up to $\pm$ 6 \AA. The unit of time is $N$ bond transposition moves.}
\label{fig:Figure_10}
\end{figure}

To study the energy behaviour in time of a sample with $N=1000$ atom, we stored the samples after every $N$ bond transposition moves. For our study we defined the unit of time as $N$  bond transpostion moves, that is one attempted bond switching move per atom. The evolution of the grain boudaries in the $1000$ atoms sample is shown in Figure \ref{fig:Figure_9}. These samples are completely flat. However, when small out-of-plane fluctuations are given, they start to buckle. Buckling caused by the grain boundaries has a long-range effect and the buckling height is also very significant (11.5 \AA) for the sample at t=100. Figure \ref{fig:Figure_10} shows the evolution of the domains in the free floating buckled graphene.

\begin{figure} 
\includegraphics[width=1.0\textwidth]{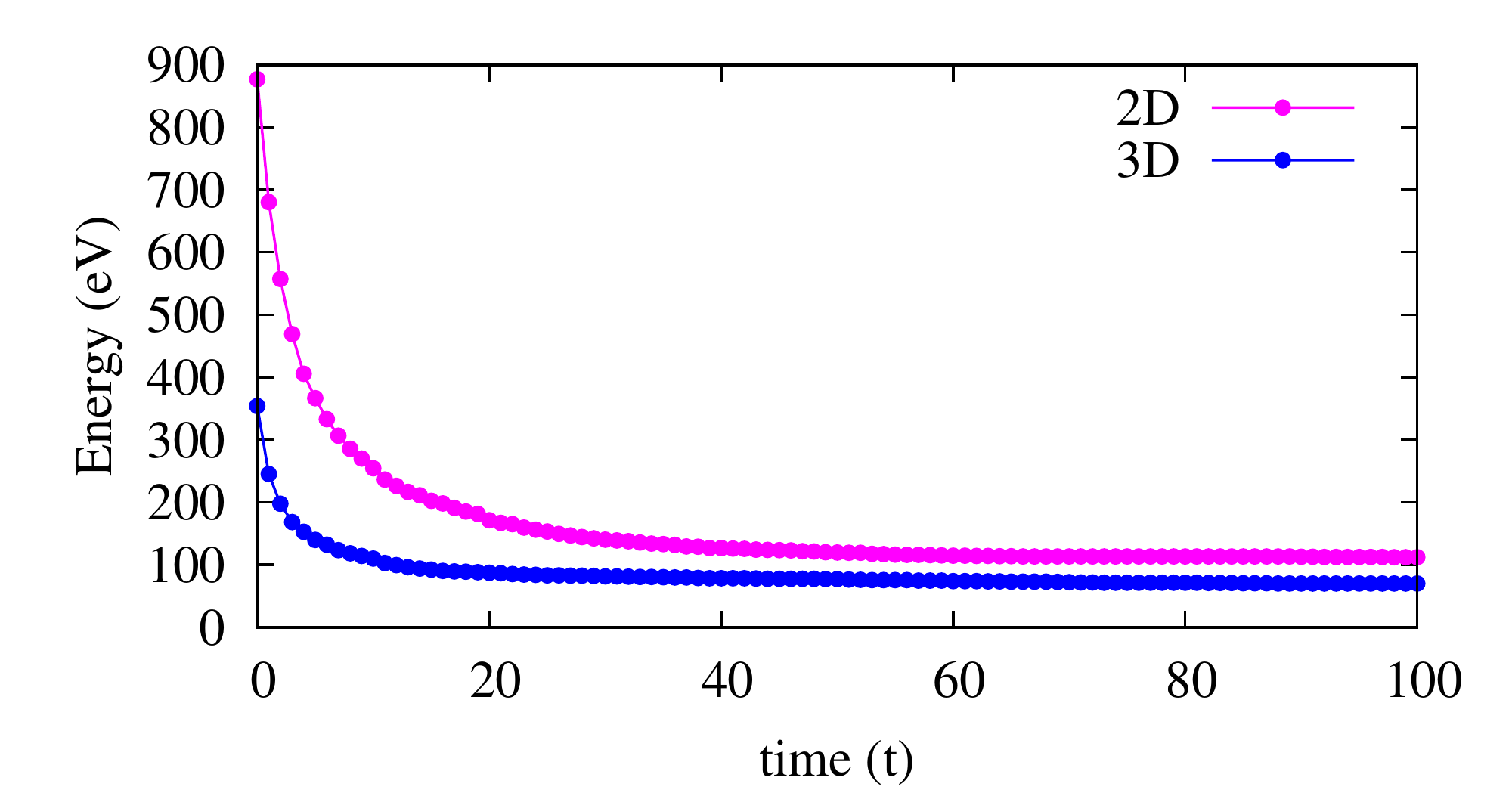}
\caption{Energy as a function of evolution time, for both the flat (planar) case and the
buckled case (with out-of-plane relaxation).} 
\label{fig:Figure_11}
\end{figure}

We study the energy evolution of domains in time for both flat (2D) and buckled modes (3D) in graphene. Our initial samples (t=0) are topologically same for both 2D and 3D cases. Figure \ref{fig:Figure_11} compares the configuration energy evolution for the 2D and 3D cases (here energy data is averaged over 5 samples). In comparison to the 2D flat case of domains, the energy converges much earlier in the 3D buckled case. The initial (t=0) buckled sample has much lower energy than the initially flat case but the energy evolution is more significant and effective in 2D in comparison to 3D case. In the process of domain evolution, domain growth becomes very slow once the system can be characterized as crystalline domains separated by grain boundaries (GBs; chains/strings of 5- and 7-fold rings). This is in agreement with experiments by Kurasch {\it {et al.}}  \cite{Kurasch2012} and Rasool {\it {et al.}} \cite{Rasool2014}. They show that continuous lines of pentagons and heptagons (GBs) are energetically stable. A review on the GBs in polycrystalline graphene by Yazyev {\it {et al.}} \cite{Yazyev2014a} is also in good agreement with our simulations (Figure 9). 

\begin{figure} 
\includegraphics[width=1.0\textwidth]{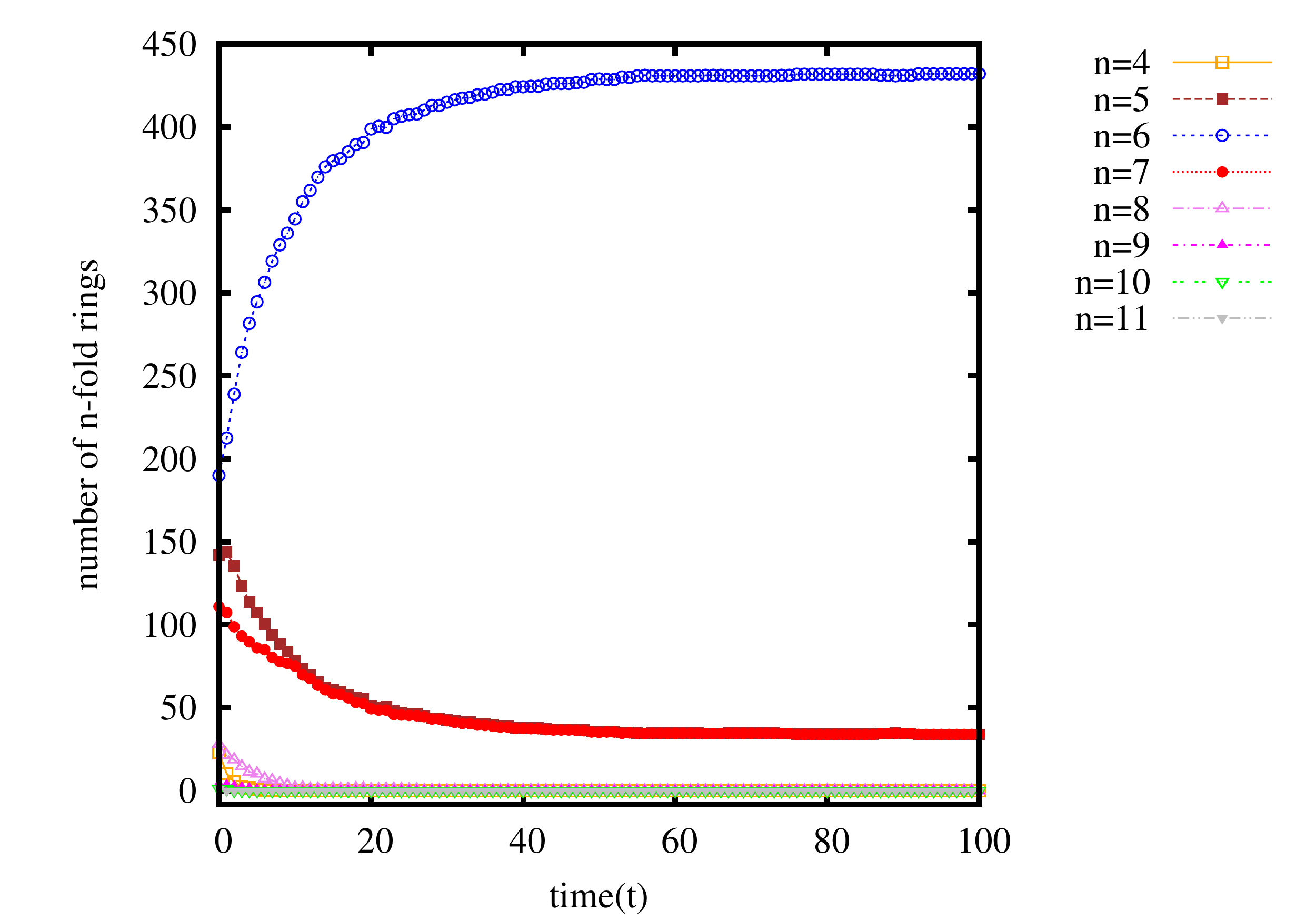}
\caption{Ring statistics as a function of evolution time, for the flat (planar) case. } 
\label{fig:Figure_12}
\end{figure}

\begin{figure} 
\includegraphics[width=1.0\textwidth]{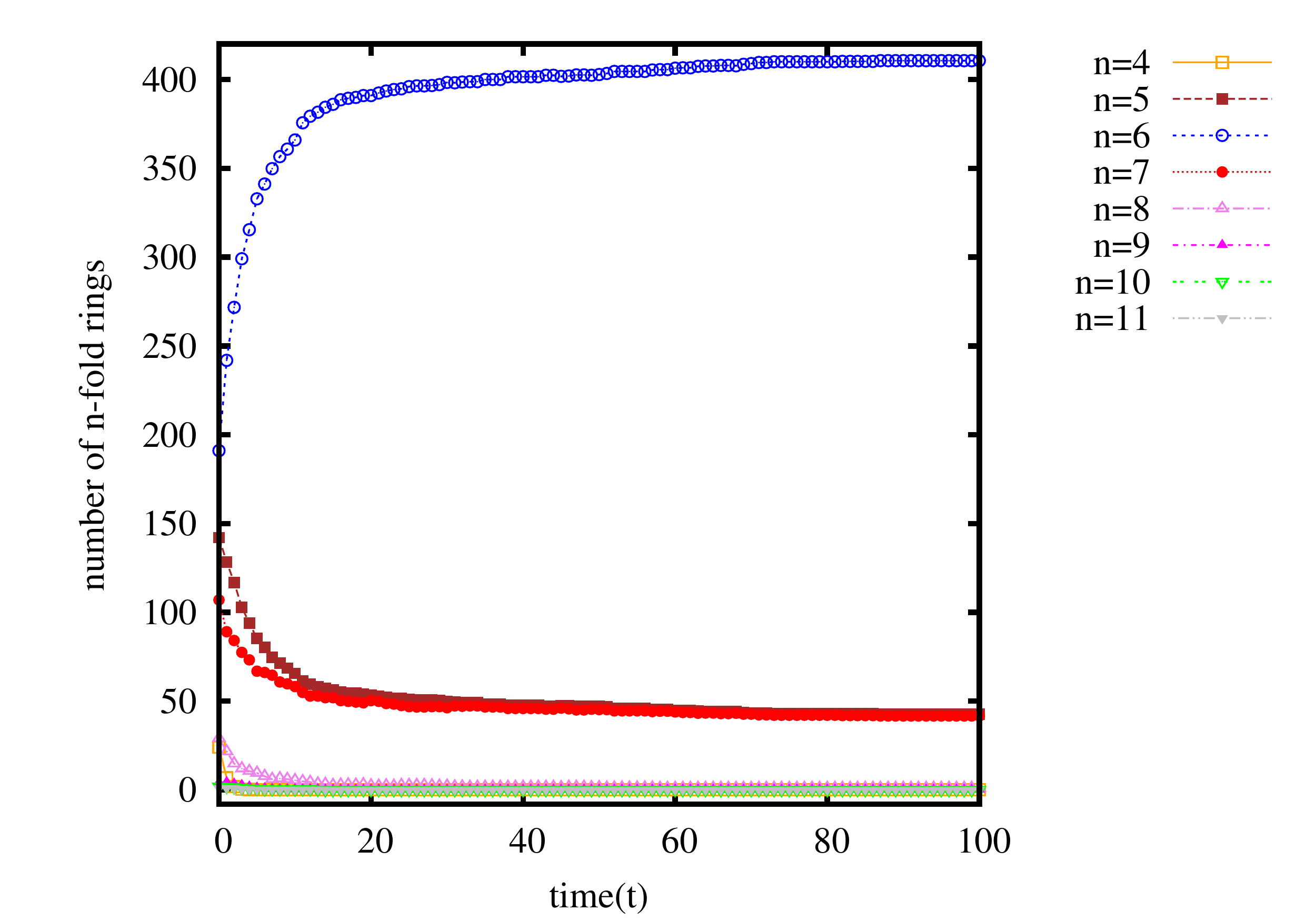}
\caption{Ring statistics as a function of evolution time, for the buckled case (with out-of-plane relaxation). } 
\label{fig:Figure_13}
\end{figure}

The evolution of different kinds of rings during the domain growth is shown in the form of ring statistics plots for both flat and buckled cases (see Figure \ref{fig:Figure_12} and Figure \ref{fig:Figure_13} respectively). During the domain evolution the number of hexagonal rings increases whereas the number of pentagon, heptogon and octagon rings decreases in order to achieve an energatically stable structure. Ring statistics is more or less same for both the cases but, after 100 units of time the number of 6-fold rings is higher in 2D (compared to 3D), which we attribute to the fact that in 2D dislocations are attracting each other; hence, they come together and are annealed during the evolution process.  At the end they form GBs which are energetically stable (which is also an experimental observation). In 3D, however, dislocations repel each other and hence they do not come together to form long grain boundaries. We have analyzed the sample topology of the 3D relaxed sample after 100 units of time and found many disclinations (single pentagon or heptagons rings) and dislocations present in the sample.

To observe the effect of substrate on the domain evolution process we introduce the confinement energy term (Eq. 9) in the potential and minimized the buckled sample (t=100) for different values of $K$ (eV \AA$^{-2}$). The buckling height decreases from 11.5 \AA ~ for $K=0$ eV \AA$^{-2}$ to 7.7 \AA ~ for $K=0.001$ eV \AA$^{-2}$, and it decreases further to 4.2 \AA ~ for $K=0.01$ eV \AA$^{-2}$. In the recent paper by Tison {\it{et al.}} \cite{Tison2014}, grain boundaries in graphene on a SiC substrate are analyzed in high resolution in three dimensions by means of scanning tunneling microscopy (STM). They find that maxima in height vary between 4 \AA ~ and 8 \AA ~ for different misorientation angles of GBs, and that the buckling in lateral dimensions extends to typically 5-20 \AA. This experimentally determined height of the grain boundaries is in excellent agreement with the height of the buckling that we obtained at different values of $K$. Thus, we infer that reasonable values of $K$ are in the range of 0.001-0.01 eV \AA$^{-2}$.

\section{CONCLUSION}

We have developed a novel potential which includes an out-of-plane buckling term which is optimized by DFT calculations. This extra term in the semi-empirical potential is essential to simulate accurately buckling in graphene. This potential uses an explicit list of the bonds between the atoms, and is therefore very cheap in terms of computational requirements. As this potential is an extension of the Kirkwood potential, bending and streching have been completely separated without interference between them. The potential reproduces very well both the structural properties and the energies of defect-free graphene, defective graphene, and graphene containing well-defined point or reconstructed defects. 
Considering the defect energies, we have revealed that calculations which do not take long-range relaxations into account, which is typically the case in literature, suffer from large systematic errors, since these relaxations can lower the formation energies even by a factor of 2 or 3. Although our potential cannot deal with extrinsic (dopant) atoms, this limitation could be overcome by a multi-method approach, where the local environment of dopant atoms is treated with e.g. DFT, while the long-range relaxations outside of this local environment are calculated using the novel semi-empirical potential. 
\\The current potential enables the accurate simulation of other defects as well, including line dislocations and grain boundaries. The experimentally well-known buckling and rippling of graphene is convincingly demonstrated for these defects. Without the out-of-plane component of the potential, these phenomena could not have been properly described. Furthermore, a long-standing paradox of the divergence in energy of a separating Stone-Wales defect has been solved: the out-of-plane energy contribution leads to stabilization of the energy at large separations.

In the future, the current potential could be used in simulations whereby interaction with supports, such as Ir(111) surfaces with steps at edges at the surface, can be simulated by hard interactions between the support and the graphene, whereas the typical buckling and rippling of the graphene can be simulated with the present potential. The current potential could also be modified for simulation of free edges in graphene, since free edges play a significant role in graphene or carbon nanotube morphologies.

Finally, we mention that the current approach can possibly be extended to the development of new potentials for other two-dimensional atomic crystals \cite{Geim2013} such as h-BN, MoS$_2$, and WSe$_2$, so that long-range structural defects can be reliably simulated in these 2D materials as well.

\begin{acknowledgement}
This project is funded by FOM-SHELL-CSER programme (12CSER049). This work is part of the research programme of the Foundation for Fundamental Research on Matter (FOM), which is part of the Netherlands Organisation for Scientific Research (NWO). MAvH acknowledges NWO for a VIDI grant. NM is supported in part by the Natural Sciences and Engineering Research Council of Canada, the Fonds de recherche du Qu\'{e}bec - Nature et technologies and the Research Chair program of Universit\'{e} de Montr\'{a}l. We thank Ingmar Swart and Vladimir Juricic (Utrecht University) for useful discussions.
\end{acknowledgement}

\bibliography{Graphene}

\end{document}